\documentstyle[psfrag,aps,multicol,epsfig,amsfonts]{revtex}

\begin{document}
\newcommand{\be}{\begin{equation}}
\newcommand{\ee}{\end  {equation}}
\newcommand{\var}{\mbox{var}}
\newcommand{\w}{\omega}
\newcommand{\rd}{{\rm d}}
\newcommand{\corr}{{\gamma}}
\newcommand{\C}{{\cal C}}
\bibliographystyle{prsty}
\widetext
\title{Statistics of selectively neutral genetic variation}
\author{A. Eriksson$^1$, B. Haubold$^2$, and B. Mehlig$^1$}
\address{\mbox{}$^1$Physics \& Engineering Physics,
        Chalmers/GU, Gothenburg, Sweden\\
         \mbox{}$^2$ LION Bioscience AG, Waldhofer Str. 98, 69123 Heidelberg,
         Germany}
\date{\today}
\maketitle{ }

\begin{abstract}
Random models of evolution are instrumental in extracting
rates of microscopic evolutionary mechanisms from empirical observations on
genetic variation in genome sequences. In this context it is necessary to know
the statistical properties of empirical observables (such as
the local homozygosity for instance). 
Previous work relies on numerical results or assumes Gaussian approximations
for the corresponding distributions. In this paper
we give an analytical derivation of
the statistical properties  of the local homozygosity and other
empirical observables assuming selective neutrality.
We find that such distributions can be very non-Gaussian.
\end{abstract}
\begin{multicols}{2}
For more than thirty years, microscopic random models of genetic
evolution have been the focus of a substantial research effort
in theoretical biology \cite{kim64,ewe79,hud90,ewe01}.
In the future, such microscopic models and their statistical analysis
will be of yet increasing significance in
this field: the amount of accurate and comprehensive data
on the genetics of viruses, bacteria  and
especially the human genome \cite{mlst,con01,ven01} has increased so considerably
that it is now possible to test microscopic models of genetic
evolution.

Genetic information is encoded in the linear sequence of nucleotides
in DNA molecules;
the four different nucleotides occurring in DNA are usually
denoted by {\tt A}, {\tt C}, {\tt G} and {\tt T}.
A sequence of a few hundred or a few thousand of these forms a gene, also referred to as a locus.
Mutations change individual nucleotides
(e.g. from {\tt A} to {\tt C}) and thus create modified versions of loci.
The resulting different types of loci are
also known as allelic types.
Because loci consist of many nucleotides
-- each of which can be changed by mutation independently from the others
-- the number of possible  allelic types is typically very large.
To a good approximation it can thus be assumed that every
mutation creates a new allelic type.
This is the defining feature of the {\em infinite-alleles model} \cite{kim64}.

Empirically, genetic variation is recorded by measuring
the frequencies $\w_a^{(l)}$ of each allelic type $a$ at each locus $l$.
Genetic variation reflects the microscopic processes of
evolutionary dynamics. The simplest model of evolution proceeds
by sampling with replacement each generation from the previous
generation (at constant population size $N$). In addition,
a number of microscopic processes take place, each
happening at a constant (but generally unknown) rate. One such process
is mutation, measured as $\theta = 2N\mu$
where $\mu$ is the probability of mutation per locus per generation
(in a haploid population).
Another such process is the exchange of genetic material between
individuals of a population measured as $C = 2Nc$
where $c$
is the probability of an exchange event per locus per generation \cite{MS94}.
$C$ is termed recombination rate.

The model of genetic evolution described here is
called the {\em constant-rate neutral mutation process},
referred to as {\em neutral process} in the following.
It is a stochastic model and assumes that no selective
forces act.
The neutral process
is one of the most significant microscopic models
of genetic evolution: not only does it provide a model for genetic variation
at loci unaffected by selection, deviations
between empirical observations and predictions of this neutral process
allow for a qualitative characterisation of selective effects
(see \cite{oht96}).

There is by now an overwhelming amount of work, both theoretical
and empirical, on the neutral process for the infinite alleles
model. A convenient way of simulating this process on a computer
is to consider genealogies of samples of a given population
\cite{kin82b,hud90,tav95} in the limit of $N\rightarrow\infty$.
Random samples are most effectively  generated by creating random
genealogies. In this way, statistics of empirical observables may
be obtained using Monte-Carlo simulations. Another possibility is
to simulate Ewen's sampling formula \cite{ewe79} which determines
the statistics of the neutral process in the limit of large $C$.
Analytical work has
mostly focused on calculating expectation values and variances of
empirical observables \cite{hud94}.  Distributions of even the
simplest empirical observables (such as the one-locus homozygosity
\cite{ewe79}) are not known analytically. The difficulty is: moments of
empirical observables are usually calculated by expanding them
into a sum of identity coefficients \cite{hud94}. This procedure is
impractical for high moments.

At the same time, the form of such distributions is of great
interest: for example, they characterise sample-to-sample fluctuations.
More importantly, they
can be used
to establish confidence intervals
for empirical observations.
To date, such confidence intervals have routinely been obtained from
Monte-Carlo calculations \cite{sou92,hau96}. Alternatively
it has been assumed that the distributions
are well approximated by Gaussians \cite{bro80}.

The aim of this paper is to calculate distributions
of empirical observables (such as
the homozygosity) in the  neutral
process for the infinite-alleles model.
The remainder is organised as follows:
first the results for a single locus are described,
and then those for two and more loci.
Finally, implications of the results are discussed.

{\em One locus.}
Consider the {\em
homozygosity} $F_2$, the probability that a pair of alleles (in a
sample of size $n$ with $m$ allelic types) has the  same allelic type. In terms
of the allelic frequencies this probability can be expressed
as (large $n$)
\be
\label{eq:defF2}
F_2 = \sum_{a=1}^m \w_a^2\,.
\ee
The statistics of $F_2$ is determined by
the moments of $F_2$
\be
    \phi_k = \langle F_2^k \rangle
\ee
where the average is over random genealogies according to the neutral process.
The $\phi_k$ may be calculated numerically in at least two ways:
by generating random genealogies \cite{kin82b,hud90,tav95}
or by evaluating Ewen's sampling formula \cite{note}.
Obtaining an analytical estimate of the $\phi_k$
is complicated by the fact that the allelic frequencies
$\w_a$ in (\ref{eq:defF2}) are not independently distributed.
 For instance, they must
 satisfy the constraint $\sum_{a=1}^m \w_a=1$.
 
To obtain analytical results we seek an approximate representation
of the neutral process in terms of {\em independent } random numbers.
When only one locus is of interest, non-recombination models apply,
irrespective of how much gene exchange actually occurs.
In this case the numbers $c_a$ of allelic types $a$ with given frequency $\w_a$
are approximately independently distributed \cite{arr94},
albeit only for sufficiently small $\w_a$.
Unfortunately this result does not yield the statistics
of $F_2$ since {\em all} frequencies $\w_a$ enter in (\ref{eq:defF2}),
and not just the small $\w_a$.

In the following we show how the distribution of $F_2$
can be determined by means of a recursion for the frequencies $\w_a$:
assume that there are $m$ allelic types with
frequencies $\w_1,\w_2,\ldots,\w_m$, obeying
the normalisation condition $ \sum_{a=1}^m \w_a=1$.
Add one allelic type; the corresponding $m\!+\!1$
frequencies $\w_a^\prime$ are defined as follows:
draw a frequency $\w_{m+1}^\prime = z_m$ with density $\Phi(z_m)$.
To ensure normalisation, define $\w_k^\prime = (1-z_m)\,\w_{k}$
for $k=1,\ldots,m$. Thus
\be
    \label{eq:pd}
    \w_a^{\prime} = z_{a-1} \prod_{b=a}^{m-1}(1-z_b)
\ee
where $z_a$ (for $a \geq 1$) are independent random variables
with density      $\Phi(z_a)$ and  $z_0=1$.
For $\Phi(z_a) = \theta\,(1-z_a)^{\theta-1}$
it follows from \cite{pat77,don86}
that (for large values of $n$)
the frequencies $\w_a$
are distributed according to the neutral
process.

The recursive definition (\ref{eq:pd}) enables us to
derive an explicit expression for the moments of $F_2$:
for large $n$
\begin{eqnarray}
    \label{eq:rec}
    F_2 &\simeq& \sum_{a=1}^m \w_a^2\,,\\
    F_2^{\prime} &\simeq& \sum_{a=1}^{m+1} {\w_a^\prime}^2 = z_m^2 + (1-z_m)^2 \sum_{a=1}^m \w_a^2 \,.
    \nonumber
\end{eqnarray}
Since
the sum on the r.h.s. does not depend on $z_m$,
it can be averaged independently from
$z_m$.
In the limit of large $n$, $F_2$ and $F_2^\prime$ have the same
distribution, $F_2\sim F_2^\prime$.
Using $\langle z^k\rangle_\Phi =
\Gamma(1+k) \Gamma(1+\theta)/\Gamma(1+k+\theta)$
and $\langle (1-z)^l\rangle_\Phi = \theta/(l+\theta)$,
\be
    \label{eq:result}
    \phi_k = \theta\,\sum_{l=0}^{k-1}{k \choose l}
     \frac{\left(2(k-l)\right)!\,\,\Gamma(2\,l+\theta)}{\Gamma(1+2\,k+\theta)}\,\phi_l\,.
\ee
Here and above $\Gamma(x)$ is the Gamma function.
Eq. (\ref{eq:result}) provides
an analytical approximation for arbitrary moments of $F_2$,
appropriate in the limit of large sample sizes $n$.

\begin{figure}
\psfrag{P(x)}[b][B]{$P(x)$}
\psfrag{x}[b][B]{$x$}
\psfrag{n50}[bl][l]{$n = 50$}
\psfrag{n500}[b][l]{$n = 500$}
\centerline{\psfig{figure=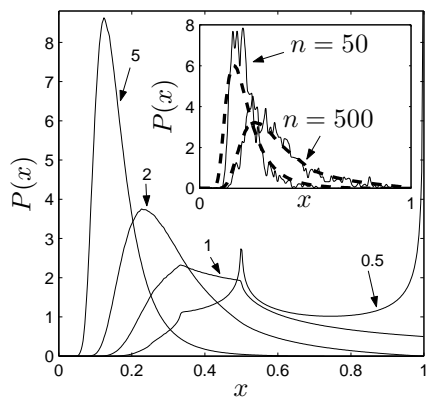,width=5.5cm}}
\vspace{3mm}
\caption{\label{fig:F2} $P(x) = \mbox{Prob}(F_2=x)$
for $L=1$ and $\theta = 0.5, 1, 2$ and $5$. Inset: analytical
results for $P(x)$ compared to the Monte-Carlo results of \protect\cite{wat77},
for $m=10$ and $n=50,500$.}
\end{figure}

One could reconstruct the distribution function
$P(x)=\mbox{Prob}(F_2=x)$ of $F_2$ from the moments
(\ref{eq:result}). It is, however, more convenient
to derive the analogue of eq. (\ref{eq:rec}) for $P(x)$ itself.
By definition [see eq. (\ref{eq:rec})],
\be
    P(x) = \int_0^1\rd z\, \Phi(z)\,P[(x-z^2)/(1-z)^2]
\ee
for $0 \leq x \leq 1$ and zero otherwise. This can be rewritten as
\be
    \label{eq:P}
    P(x) = \int_0^1 \rd y\,Q(x,y)\,P(y)
\ee
with the kernel
\be
  \label{eq:Q}
  \begin{array}{l}
      Q(x,y) = \\[1mm]
        \quad \frac{\theta}{2 a}
        \left[
            \left(\frac{1 + a}{1 + y}\right)^{\theta-1}\!\!\!\!\!\!
            \raisebox{-1mm}{+}\;
            \left(\frac{1 - a}{1 + y}\right)^{\theta-1}\!\!\!\!\!\!
            \raisebox{-1mm}{$\mbox{H}(y-x)$}\,
        \right]
        \raisebox{-1mm}{$\mbox{H}(\frac{x}{1-x}-y)$}
  \end{array}
\ee
where $a \equiv a(x,y) = \sqrt{x-(1-x)y}$
and $\mbox{H}(z)$ is the Heaviside step function. Note
that $Q(x,y)$ exhibits a divergence as $x,y \rightarrow 0$. Eq.
(\ref{eq:P}) is solved by expanding $P(x)$ in a suitable set of
basis functions on the interval $[0,1]$, resulting in an eigenvalue problem.
Fig. \ref{fig:F2} shows the resulting distributions $P(x)$  for
four values of $\theta$. Clearly the statistics of $F_2$
is very non-Gaussian.

The calculations summarised above are not only of interest
in the case of one locus, as the following paragraphs show
(in the following $L$ denotes the number of loci).

\begin{figure}
\psfrag{YLABEL}[b][B]{$P(x)$}
\psfrag{XLABEL}[b][B]{$x$}
\centerline{\psfig{figure=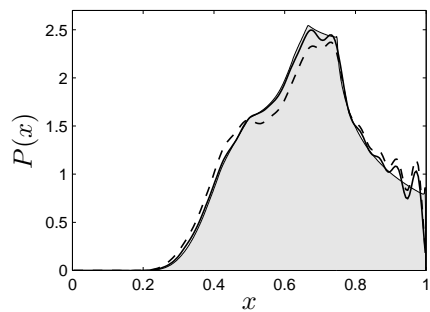,width=5.5cm}}
\vspace{3mm}
\caption{\label{fig:L2}
$P(x)$\,$=$\,$\mbox{Prob}(F_2=x)$ for two loci ($L=2$)
and $\theta = 0.5$, in the limit of large $C$ and $n$ (shaded).
Also shown are results of Monte-Carlo simulations for $n=100$ and
$C=10$ (solid line) and $C=1$ (dashed line).}
\end{figure}

{\em Two loci}. In the case of two loci ($L=2$) on
the same stretch of DNA, the joint distribution of allelic
frequencies $\w^{(l)}_a$ depends on the rate $C$ of gene exchange.
Consider (for large $n$)
\be
    \label{eq:defF2L}
    F_2 = \frac{1}{L}\sum_{l=1}^LF_2^{(l)}\,,\quad
    F_2^{(l)} = \sum_a {\omega_a^{(l)}}^2\,.
\ee
In the limit of large $C$, the two genealogies
for $l=1$ and $l=2$ are essentially independent and
the frequencies $\w^{(l)}_a$ are well approximated by
(\ref{eq:pd}) for each $l$ (and large $n$).
The distribution
$P(x) = \mbox{Prob}(F_2=x)$ is thus obtained
from the single-locus $P(x)$ by convolution.
The resulting distribution is shown in Fig. \ref{fig:L2}.
Empirically determined recombination rates
are often so large that this result for $P(x)$
is a good approximation: in Fig. \ref{fig:L2} two distributions
of $F_2$ are shown, for $n=100$, $\theta=1/2$ and $C=1$ and $10$, obtained
from Monte-Carlo simulations.
One observes good agreement with the prediction (shaded), even
for values of $C$ as low as $C=1$.
It must be emphasised that the distribution is markedly non-Gaussian.
The wiggles in the Monte-Carlo results are statistically significant;
they are a consequence of the finite sample size ($n=100$).

{\em Many loci}. When $L\gg 1$, and in the limit of large $C$, the
distribution of $F_2$ [as defined in (\ref{eq:defF2L})] is Gaussian, and its moments are obtained as
\be
    \phi_k = \left[ 1 + {k \choose 2}\frac{2\,\theta}{(2 + \theta)(3 +
    \theta)}\frac{1}{L}\, \right] (1 + \theta)^{-k}\,.
\ee

\begin{figure}
\psfrag{y}[b][][1.2]{$z_a$}

\psfrag{a}{$a$}

\psfrag{i}{$b$}

\psfrag{j}{$a$}

\psfrag{x}{$x$}

\psfrag{(a)}{(a)}

\psfrag{(b)}{(b)}

\centerline{\psfig{figure=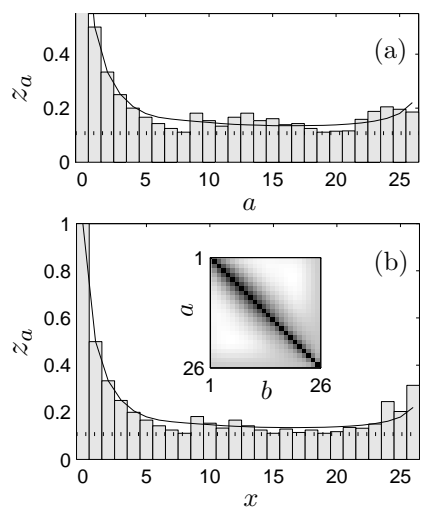,width=5.5cm}}

\vspace{3mm} \caption{\label{fig:neutral}
(a) frequencies $z_a$ from empirical $\w_a$
(locus {\em GltA} in {\em C. jejuni} \protect\cite{din01}),
compared to the neutral model for $n\rightarrow\infty$
(dashed line). Also shown are results of Monte-Carlo
simulations for finite $n=194$ (solid line). (b) is a
similar plot with data taken from one Monte-Carlo sample.
The inset shows the 
correlation strength between $z_a$ and $z_b$ for $n=194$ and 27
alleles. Black corresponds to full correlation. }
\end{figure}

{\em Discussion.} 
In an empirical data set, $n$ (and $m$) are necessarily finite. It must then 
be asked: 
to which extent are the
$z_a$ independently and identically distributed for finite $n$ (and $m$)?
Fig. \ref{fig:neutral}(a) shows $z_a$-values determined
from empirical data on {\em C. jejuni} \cite{din01},
at the locus {\em GltA} ($n=194$ and $m=27$), 
in comparison with 
the theory for $n=\infty$. 
The empirical $z_a$ are approximately identically
distributed, except at the edges where finite-size effects are observed
(remember that $z_0\equiv 1$). 
Monte-Carlo simulations for $n=194$ and $m=27$ confirm the
effect of finite sample size.
Fig. \ref{fig:neutral}(b) is a similar plot with
data taken from one Monte-Carlo sample. 
The inset of Fig. \ref{fig:neutral}(b) shows that
that the $z_a$ are indeed independently distributed.
It can be concluded that
the theory works well in the present case.

In the remainder two implications of our results are discussed.  
First, in practice it is necessary to decide whether empirically
 observed frequencies at a given locus are consistent
 with the neutral process. The standard
 statistical test (see \cite{ewe79} p. 263) uses the distribution
 of $F_2$ as an input (albeit with the number $m$ of allelic types
 as a parameter and not $\theta$ as in the above equations).
 Since the distribution of $F_2$ was unknown, it was
 usually determined by Monte-Carlo simulations. Now, however,
 the result (\ref{eq:P},\ref{eq:Q}) can be used:
 for
 $m\, {\scriptstyle\gtrsim}\, \log\,n$,
 eqs. (\ref{eq:P},\ref{eq:Q})
 apply independently of whether $m$ or $\theta$ is
 taken as the parameter. The corresponding distributions
 are compared to Monte-Carlo data \cite{wat77} in Fig. \ref{fig:F2}.
 Shown are two cases: $m=10,n=50$ and $m=10,n=500$.
 In both cases, the agreement between our results and
 those of Monte-Carlo simulations is very good.

Second, many recent empirical studies (see for instance
\cite{sou92,hau96,mai98}) have analysed
the extent of gene exchange. A common measure is
the variance $V_{\rm D}$ of the number
of pairwise differences at all loci under consideration.
In the limit of $C\rightarrow\infty$ (linkage equilibrium)
$\langle V_{\rm D} \rangle = \langle \sum_{l=1}^L(1-F_2^{(l)})F_2^{(l)}\rangle$
(for the neutral process this evaluates to
to $L\theta(4+\theta)/[(1+\theta)(2+\theta)(3+\theta)]$, see
\cite{hud94}).
However for finite values of $C$ (linkage disequilibrium),
and especially for small $C$,
the expected value of $V_{\rm D}$ is larger.
The empirically determined value of $V_{\rm D}$ can be compared to a critical
value obtained under the null hypothesis that all loci
are in linkage equilibrium. The corresponding null distribution
is usually obtained using Monte-Carlo simulations \cite{sou92,hau96}.

\begin{figure}
\psfrag{PVE(x)}[b][B]{$\mbox{Prob}(V_D = x)$}
\psfrag{x}[b][B]{$x$}
\centerline{\psfig{figure=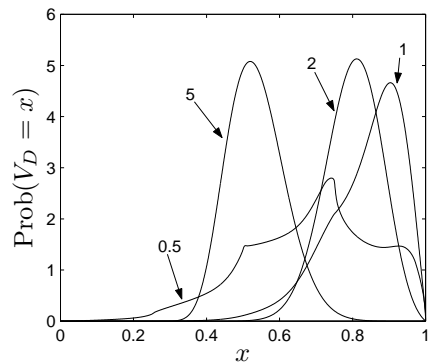,width=5.5cm}}
\vspace{3mm}
\caption{\label{fig:VD} Null distribution of $V_{\rm D}$
for $L=4$, and $\theta = 0.5,1,2$ and $5$ (in the limit of
large $C$, the range of $V_{\rm D}$ is $0 \leq V_{\rm D} \leq L/4$).}
\end{figure}

In cases where the neutral model applies, the null distribution
of $V_{\rm D}$ can be determined from
eqs. (\ref{eq:result}), (\ref{eq:P}) and (\ref{eq:Q}).
Consider first the case of large $L$, where the null distribution
is approximately Gaussian. Using
$V_{\rm D} \sim \sum_{l=1}^L(1-F_2^{(l)})F_2^{(l)}$
for large $C$, one obtains 
\begin{eqnarray}
    \label{eq:varvd}
    &&\mbox{Var}\left[\, V_{\rm D}\, \right] =
    L\, \left[\phi_4 - 2\, \phi_3 + \phi_2 - (\phi_1 - \phi_2)^2\right]\\
    &=&L \frac{2\theta(1872\!-\!420\,\theta\!-\!584\,\theta^2\!+\!229\,\theta^3\!+\!1
63\,\theta^4\!+\!23\,\theta^5\!+\!\theta^6)}{(1+\theta)^2 (2+\theta)^2 (3+\theta)
^2 (4+\theta) (5+\theta) (6+\theta) (7+\theta)}\,.\nonumber
\end{eqnarray}
This variance is always larger than the corresponding quantity
in a random shuffling scheme  \cite{sou92,hau96} because the latter
is conditioned on the homozygosity, and not on $\theta$.
When $L$ is small, the null distribution will be very non-Gaussian,
as the above results for the distribution of $F_2$ show.
In Fig. \ref{fig:VD}, the null distribution of $V_{\rm D}$
[as determined from (\ref{eq:P},\ref{eq:Q})]
is shown for the case of $L=4$ and for four values of $\theta$.
Note that the forms of the distributions imply
large, asymmetric confidence intervals.
Finally, for
$m\, {\scriptstyle\gtrsim}\, \log\,n$,
the distributions
in Fig. \ref{fig:VD} are insensitive to whether
the process is conditioned on fixed $\theta$ or fixed $k$
\cite{note2}.

{\em Conclusions.}
We have shown that distribution functions of empirical
observables measuring genetic
diversity in selectively neutral populations
may exhibit strong non-Gaussian tails. We have
found analytical approximations for these
distributions, valid for large sample sizes and in the limit where
gene exchange is frequent; and have discussed implications
for the statistical analysis of genetic variation.
It is highly desirable to extend
the present results to the case where gene exchange
is rare, corresponding to clonal or nearly clonal populations.

\narrowtext

\end{multicols}

\begin{thebibliography}{10}
\bibitem{note2}
Note, however, that for
small $\theta$ any error in its estimation may influence the
distribution of $V_{D}$, see Fig. \ref{fig:VD}.
\end{thebibliography}
\end{document}